\documentstyle[sprocl,psfig,amssymb]{article}

\def\be{\begin{equation}}
\def\ee{\end{equation}}
\def\bea{\begin{eqnarray}}
\def\eea{\end{eqnarray}}

\begin{document}

\title{KADANOFF--BAYM EQUATIONS WITH INITIAL CORRELATIONS}

\author{D. KREMP, D. SEMKAT, M. BONITZ}

\address{Universit\"at Rostock, Fachbereich Physik, Universit\"atsplatz 3,\\
D-18051 Rostock, Germany} 


\maketitle\abstracts{
The Kadanoff--Baym equations (KBE) are usually derived
under the assumption of the weakening of initial correlations (Bogolyubov's
condition) and, therefore, fail to correctly describe the short time behavior.
We demonstrate that this assumption is not necessary. Using functional
derivatives techniques, we present a straightforward
generalization of the KBE which allows to include arbitrary initial correlations
and which is more general than previous derivations. As a result, an additional
collision integral is obtained which is being damped out after a few
collisions. Our results are complemented with numerical investigations
showing the effect of initial correlations.
}


\section{Introduction and basic equations}\label{intro}

Nonequilibrium properties of many-particle systems are traditionally described
by kinetic equations of the Boltzmann type. Despite their fundamental
character, these equations have well-known principal shortcomings, e.g.
(i) the short--time behavior ($t<\tau_{corr}$) cannot be described correctly,
(ii) the kinetic or the quasiparticle energy is conserved only instead of the
total (sum of kinetic and potential) energy,
(iii) no bound states are contained, and
(iv) the equations yield the equilibrium distribution and thermodynamics for
ideal particles.

Since the 1950ies, there were many efforts to overcome these deficiencies.
In the time--diagonal formalism, Prigogine and Resibois \cite{PrigRes}, Zwanzig
\cite{Zwanzig}, and Zubarev \cite{Zubarev} derived exact kinetic equations which
(i) are nonlocal with respect to the time (non--Markovian), (ii) contain an
initial correlation contribution, and (iii) yield the correct conservation laws
for nonideal systems.
On the other hand, Kadanoff and Baym \cite{KadBaym}, and Keldysh \cite{Keldysh}
developed equations for the two--time correlation functions. These
well--known Kadanoff--Baym equations (KBE) have a non--Markovian character, too.
The structure of approximations is determined by a single central quantity, the
self--energy, and conservation laws for nonideal systems are fulfilled for any
conserving approximation for the self--energy \cite{KB61}. However, the original
KBE contain no contribution of initial correlations and so do not reach the
degree of generality e.g. of the equation of Prigogine with respect to
this feature.

In order to solve this problem, several methods have been used, e.g.
analytical continuation to real times which allows to include equilibrium
initial correlations \cite{KadBaym,Dan84_1,Wagner,MR}, and perturbation theory
with initial correlations \cite{Fujita,Hall,Dan84_1}.
A convincing solution has been presented by Danielewicz \cite{Dan84_1}.
He developed a perturbation theory for a general initial
state and derived generalized KBE which take into account arbitrary
initial correlations.
A straightforward and very intuitive method which is not based on perturbation
theory \cite{KSB85,SKB99}, uses the equations of motion for the Green's
functions.

Starting point of this non-perturbative method is the Martin--Schwinger
hierarchy \cite{MS59}, a system of coupled equations for the
$s$--particle Green's functions defined on the Keldysh--Schwinger contour
${\cal C}$ by
\begin{equation}\label{gskontur}
g_{1...s}(1...s,1'...s')=\left(\frac{1}{i}\right)^s
\left\langle T_{{\cal C}}\left[\Psi(1)...\Psi(s)\Psi^+(s')...\Psi^+(1')\right]
\right\rangle,
\end{equation}
where $\Psi$ is the field operator, $T_{{\cal C}}$ the time ordering operator on
the contour, and $\langle ...\rangle$ denotes averaging over the density
operator $\rho$. The first hierarchy equation is the equation of motion
for the single--particle Green's function
\begin{eqnarray}\label{MSH1}
&&\int\limits_{{\cal C}} d{\bar 1}\left\{g_1^{0^{-1}}(1,{\bar 1})-U(1,{\bar 1})
\right\}g_1({\bar 1},1';U)\nonumber\\
&=&\delta (1-1')\pm i\int d2\,V(1-2)g_{12}(12,1'2^+),
\end{eqnarray}
with $U$ being an external potential and $g_1^{0^{-1}}$ the inverse Green's
function
\begin{equation}
g_1^{0^{-1}}(1,1')=
\left(i\frac{\partial}{\partial t_1}+\frac{\nabla_1^2}{2m}\right)\delta (1-1').
\end{equation}
Eq. (\ref{MSH1}) is not a closed equation for $g_1(1,1')$, because the
interaction leads to a coupling of the one--particle to the two--particle
function and so on.

To find a closed equation for the one--particle Green's function, one has to
consider three main steps:
\begin{itemize}
\item[I] The hierarchy can be decoupled formally by introduction of the
self--energy, taking into account, that $g_{12}$ can be derived from $g_1$ by
means of functional derivation,
\begin{eqnarray}\label{Sigmadef}
&&\int\limits_{{\cal C}} d{\bar 1}\,\Sigma(1,{\bar 1})g_1({\bar 1},1')
=\pm i\int d2\,V(1-2)g_{12}(12,1'2^+)\nonumber\\
&=&\pm i\int d2\,V(1-2)
\left\{\pm\frac{\delta g_1(1,1';U)}{\delta U(2^+,2)}
+g_1(1,1')g_1(2,2^+)\right\}.
\end{eqnarray}
\item[II] An important consistency criterion for $\Sigma$ is the correct limit
for $t\to t_0$. From Eq. (\ref{Sigmadef}), it is obvious that
\begin{eqnarray}\label{Sigmalim}
&&\lim_{t_1=t'_1\to t_0}\int\limits_{{\cal C}} d{\bar 1}\,
\Sigma(1,{\bar 1})g_1({\bar 1},1')
=\pm i\int d{\bf r}_2\,V({\bf r}_1-{\bf r}_2)
\left[c({\bf r}_1{\bf r}_2,{\bf r}'_1,{\bf r}'_2;t_0)\right.\nonumber\\
&&+\left.g_1({\bf r}_1{\bf r}'_1,t_0)g_1({\bf r}_2{\bf r}'_2,t_0)
\pm g_1({\bf r}_1{\bf r}'_2,t_0)g_1({\bf r}_2{\bf r}'_1,t_0)
\right],
\end{eqnarray}
where $c({\bf r}_1{\bf r}_2,{\bf r}'_1{\bf r}'_2;t_0)$ is the initial
binary correlation.
Therefore, $\Sigma$ must contain, in addition to the Hartree--Fock
contribution, another time--local part $\Sigma^{in}$,
\begin{eqnarray}\label{Sigmastruc}
\Sigma(1,1')&=&\Sigma^{HF}(1,1')+\Sigma^c(1,1')+\Sigma^{in}(1,1'),
\nonumber\\
\Sigma^{in}(1,1')&=&\Sigma^{in}(1,{\bf r}'_1t_0)\delta(t'_1-t_0),
\end{eqnarray}
and analogous for the adjoint equation,
\begin{eqnarray}\label{Sigmastrucadj}
{\hat\Sigma}(1,1')&=&\Sigma^{HF}(1,1')+\Sigma^c(1,1')+\Sigma_{in}(1,1'),
\nonumber\\
\Sigma_{in}(1,1')&=&\Sigma_{in}({\bf r}_1t_0,1')\delta(t_1-t_0).
\end{eqnarray}

\item[III] In order to make $\Sigma$ unique, initial (or boundary) conditions
are necessary. For example, in thermodynamic equilibrium, the
solution is uniquely fixed by the Kubo--Martin--Schwinger condition.
Furthermore, in the nonequilibrium case, we can use
the familiar Bogolyubov condition of weakening of initial correlations,
\begin{equation}\label{wic}
\lim\limits_{t_0\to -\infty}g_{12}(12,1'2')|_{t_0}=
[g_1(1,1')g_1(2,2')\pm g_1(1,2')g_1(2,1')]|_{t_0}.
\end{equation}
In this case, $\Sigma^{in}=0$, and we recover the original KBE \cite{KSB85}.

The most general initial condition has the form
\begin{eqnarray}\label{g12_0}
\lim_{t_1=t_2=t'_1=t'_2\to t_0}g_{12}(12,1'2')&=&
[g_1(1,1')g_1(2,2')\pm g_1(1,2')g_1(2,1')]|_{t_0}\nonumber\\
&&+c({\bf r}_1{\bf r}_2,{\bf r}'_1{\bf r}'_2;t_0).
\end{eqnarray}
\end{itemize}

\section{Self--energy and initial correlations}\label{secsigma}
In order to determine the self--energy, we start from Eq. (\ref{Sigmadef}).
With this definition we have obtained a formally closed
equation for the one--particle Green's function on the contour
which may be cast into the form
\begin{equation}\label{Dyson}
\int\limits_{{\cal C}} d{\bar 1}\left\{g_1^{0^{-1}}(1,{\bar 1})-U(1,{\bar 1})
-\Sigma(1,{\bar 1})\right\}g_1({\bar 1},1',U)=\delta(1-1').
\end{equation}
This equation is a compact notation of the Kadanoff--Baym equations and is
sometimes called Dyson--Schwinger equation.
The same procedure for the adjoint of equation
(\ref{MSH1}) leads to the adjoint Dyson equation with the self--energy
${\hat\Sigma}$. Eqs.(\ref{Sigmastruc},\ref{Sigmastrucadj}) show that
$\Sigma={\hat\Sigma}$ for all times $t,t'>t_0$.

Further analysis of Eq. (\ref{Sigmadef}) requires to
evaluate the functional derivative $\delta g/\delta U$, for which a simple
procedure has been given \cite{KB61}.

Starting from the Dyson equation, which, for $t,t' > t_0$, can be written in the
form
\begin{equation}\label{g^-1g}
\int\limits_{{\cal C}} d{\bar 1}\,g_1^{-1}(1,{\bar 1})g_1({\bar 1},1')=
\delta (1-1'),
\end{equation}
where we introduced the inverse Green's function
\begin{equation}\label{g^-1_l}
g_1^{-1}(1,1')=g_1^{0^{-1}}(1,1')-U(1,1')-\Sigma(1,1'),
\end{equation}
one obtains easily by functional differentiation
\begin{equation}\label{Abl. g-1g=1}
\int\limits_{{\cal C}} d{\bar 1}\,\frac{\delta g_1^{-1}(1,{\bar 1})}
{\delta U(2',2)}g_1({\bar 1},1')=
-\int\limits_{{\cal C}} d{\bar 1}\,g_1^{-1}(1,{\bar 1})
\frac{\delta g_1({\bar 1},1')}{\delta U(2',2)}.
\end{equation}
Using (\ref{g^-1_l}), the general solution of this equation and its adjoint is
found to be
\begin{eqnarray}\label{L_mit_Sig}
\frac{\delta g_1(1,1')}{\delta U(2',2)}=g_1(1,2')g_1(2,1')
&&+\int\limits_{{\cal C}} d{\bar 1}d\overline{\overline{1}}\,g_1(1,{\bar 1})
\frac{\delta\left[\Sigma({\bar 1},\overline{\overline{1}})+
\Sigma_{in}({\bar 1},\overline{\overline{1}})\right]}{\delta U(2',2)}
g_1(\overline{\overline{1}},1')\nonumber\\
&&\pm C(12,1'2')\, =\, \pm L(12,1'2'),
\end{eqnarray}
where $C$ is an arbitrary function which obeys the homogeneous equation,
i.e.
\begin{equation}\label{g^-1c}
\int\limits_{{\cal C}} d{\bar 1}\,g_1^{-1}(1,{\bar 1})C({\bar 1}2,1'2')=0.
\end{equation}
There are three similar conditions, one following from the crossing symmetry
($1 \leftrightarrow 2$) and two from the adjoint Dyson equation.

Let us now analyze the physical and mathematical consequences of the
function $C(12,1'2')$. To this end, we consider Eq. (\ref{L_mit_Sig})
in the limit $t,t'\to t_0$. In this case, the integral over the
contour vanishes, and it directly follows
\begin{equation}\label{AB_c}
L(r_1,r_2,r'_1,r'_2,t_0)=c(r_1,r_2,r'_1,r'_2,t_0)\pm
g_1(r_1,r'_2,t_0)g_1(r_2,r'_1,t_0).
\end{equation}
Hence, the function $c(t_0)$ is to be identified with initial binary
correlations.

In order to explore the temporal evolution of the function $C(12,1'2')$, we
solve Eq. (\ref{g^-1c}) (and the three analogous relations) together with the
initial condition
\begin{equation}\label{c0}
C(12,1'2')|_{t_1=t_2=t'_1=t'_2=t_0}=C(t_0).
\end{equation}
The result is
\begin{eqnarray}\label{Ckeldysh}
C(12,1'2')&=&
\int\limits_{{\cal C}} d{\bar 1}d{\bar 2}d\overline{\overline{1}}
d\overline{\overline{2}}\,
g_1(1,{\bar 1})g_1(2,{\bar 2})\,
c({\bar 1}{\bar 2},\overline{\overline{1}}\overline{\overline{2}})\,
g_1(\overline{\overline{2}},2')g_1(\overline{\overline{1}},1'),\\
\label{cdelta}
\mbox{with}\quad
c({\bar 1}{\bar 2},\overline{\overline{1}}\overline{\overline{2}})&=&
c({\bar r}_1t_0,{\bar r}_2t_0,\overline{\overline{r}}_1t_0,
\overline{\overline{r}}_2t_0)\nonumber\\
&&\times\delta({\bar t}_1-t_0)\delta({\bar t}_2-t_0)
\delta(\overline{\overline{t}}_1-t_0)\delta(\overline{\overline{t}}_2-t_0).
\end{eqnarray}

Let us now come back to the self--energy. Introducing Eqs. (\ref{L_mit_Sig})
and (\ref{Ckeldysh}) into (\ref{Sigmadef}), we find the following functional
equation for $\Sigma$:
\begin{eqnarray}\label{Sigmaexpl}
\Sigma(1,1')&=&\pm i\int d2\,
V(1-2)\left\{\pm\int\limits_{{\cal C}} d{\bar 1}\,g_1(1,{\bar 1})
\frac{\delta\left[\Sigma({\bar 1},1')+\Sigma_{in}({\bar 1},1')\right]}
{\delta U(2^+,2)}\right.\nonumber\\
& &\hspace{2cm}+\delta(1-1')g_1(2,2^+)\pm \delta(2-1')g_1(1,2^+)
\nonumber\\
& &\hspace{2cm}
\left.+\int\limits_{{\cal C}} d{\bar 1}d{\bar 2}d\overline{\overline{2}}\,
g_1(1,{\bar 1})g_1(2,{\bar 2})\,
c({\bar 1}{\bar 2},1'\overline{\overline{2}})\,
g_1(\overline{\overline{2}},2^+)\right\}
\end{eqnarray}
An analogous equation follows readily for ${\hat\Sigma}$.
With Eq. (\ref{Sigmaexpl}), the self--energy is given as a functional of the
interaction, the initial correlations and the one--particle Green's function.
From the definition of  $c$ , Eq. (\ref{cdelta}), it is obvious that the last
contribution on the r.h.s.
is local in time with a $\delta$-type  singularity at $t=t'$. Additional terms
of this structure arise from the functional derivative. A further important
property of the self--energy follows from comparing $\Sigma$, Eq.
(\ref{Sigmaexpl}) with the corresponding expression for  ${\hat\Sigma}$. One
verifies that $\Sigma={\hat\Sigma}$ for all times $t,t'>t_0$, which means, in
particular, that for these times, a well defined inverse Green's function does
exist.

Eq. (\ref{Sigmaexpl}) is well suited  to come to approximations for the
self--energy. By iteration, a perturbation series for
$\Sigma$ in terms of  $g, V$ and $C$ can be derived which begins with
\begin{eqnarray}\label{Sigma1}
\Sigma^1(1,1')&=&\pm i\delta(1-1')\int d2\,V(1-2)g_1(2,2^+)
\nonumber\\
& &\pm i\int d2V(1-2)\int\limits_{{\cal C}} d{\bar 1}d{\bar 2}
d\overline{\overline{2}}\,g_1(1,{\bar 1})g_1(2,{\bar 2})\,
c({\bar 1}{\bar 2},1'\overline{\overline{2}})\,
g_1(\overline{\overline{2}},2^+)\nonumber\\
& &+ {\rm exchange}.
\end{eqnarray}
It is instructive, to represent formula (\ref{Sigma1}) in terms of Feynman
diagrams: 
\begin{eqnarray*}\label{Sigma1diag}
\centerline{ 
\psfig{file=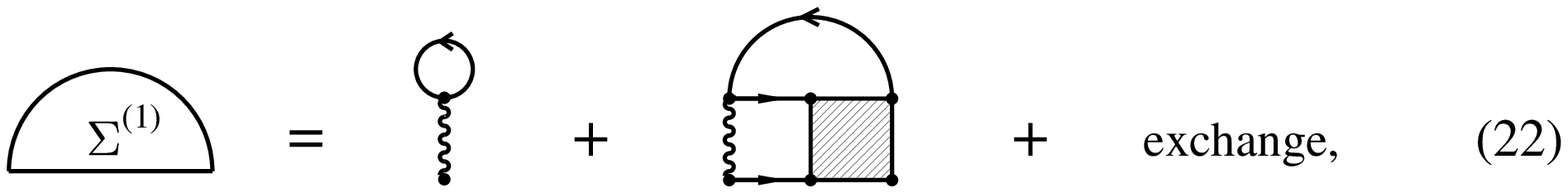,height=1.5cm,width=12cm}}
\end{eqnarray*}
where, in contrast to conventional diagram techniques, we have introduced the
initial correlation as a new basic element, drawn as a shaded rectangle.
Second order contributions are evaluated straightforwardly too, with the result
\begin{eqnarray*}\label{Sigma2diag}
\centerline{ 
\psfig{file=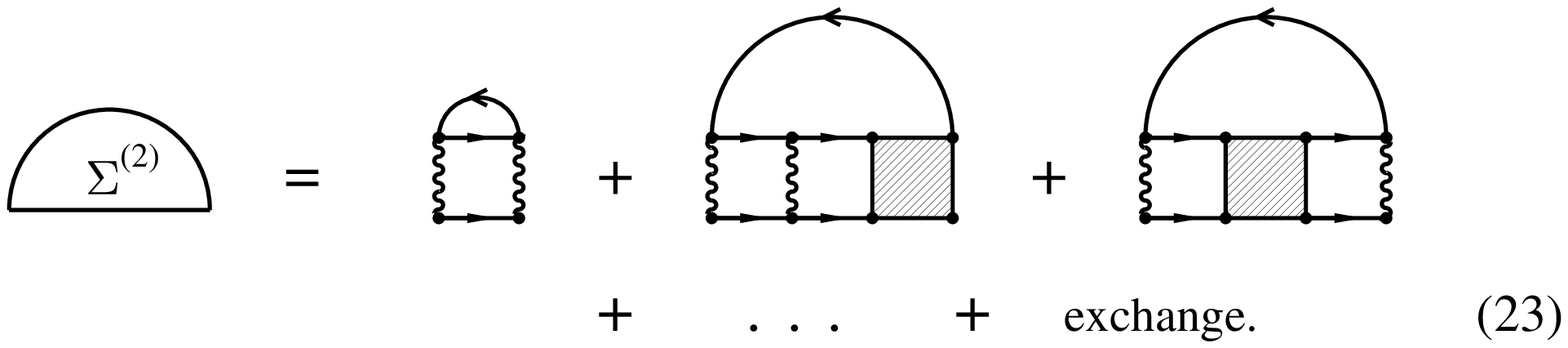,height=2.6666cm,width=12cm}}
\end{eqnarray*}
\addtocounter{equation}{2}
The analysis of the iteration scheme allows us to conclude that all
contributions to the self--energy (all diagrams) fall into two classes:
(I) the terms $\Sigma^{HF}$ und $\Sigma^{c}$ which begin and end with a
potential and (II),  $\Sigma^{in}$ - those which begin with a
potential, but end with an initial correlation. This conclusion verifies
the structure of the self--energy, Eqs. (\ref{Sigmastruc},\ref{Sigmastrucadj}),
as we discussed in the previous section.
The initial correlation part of the self--energy in fact turns out to be
temporally local (similar to the Hartree--Fock term) which is
nonzero only if $t'_1$ (or, in the adjoint case, $t_1$) is equal to the initial
time $t_0$.
The same result was obtained by Danielewicz based on his
perturbation theory for general initial states \cite{Dan84_1}.

If one considers the first two iterations for the self--energy, Eqs.
(\ref{Sigma1diag},\ref{Sigma2diag}), more in detail, it becomes evident that, in
the initial correlation contribution, in front of the function $c$, appear just
the ladder terms which lead to the buildup of the two--particle Green's
function. Thus, obviously, the iteration ``upgrades'' the product of retarded
one--particle propagators in the function $C$ to a full two--particle
propagator, in the respective order, i.e. $\Sigma^{in}$ is of the form
\begin{eqnarray}\label{Sig_in}
\Sigma^{in}(11')&=&\pm i\int d2 \,V(1-2)\int d{\bar r}_1d{\bar r}_2
d\overline{\overline{r}}_1d\overline{\overline{r}}_2
\nonumber\\
& \times &g_{12}^R(12,{\bar r}_1t_0,{\bar r}_2t_0)
c({\bar r}_1t_0,{\bar r}_2t_0,1',
\overline{\overline{r}}_2t_0)
g_1^A(\overline{\overline{r}}_2t_0,2^+)\delta(t'_1-t_0).
\end{eqnarray}
Notice especially, that this renormalization occurs not symmetrically, i.e.
in the adjoint of Eq. (\ref{Sig_in}),
$\Sigma_{in}\equiv\left[\Sigma^{in}\right]^{\dagger}$ appears the adjoint
propagator $g_{12}^A$.

\section{Generalized Kadanoff--Baym equations}\label{GKBE}
Let us now come back to the Kadanoff--Baym equations. In order to discuss the
influence of initial correlations, we insert expression (\ref{Sigmastruc}) into
Eq. (\ref{MSH1}),
\begin{eqnarray}\label{kbecont}
&&\int\limits_{{\cal C}}d{\bar 1}\left[g_1^{0^{-1}}(1{\bar 1})-U(1{\bar 1})-
\Sigma^{HF}(1{\bar 1})\right]g_1({\bar 1}1')\nonumber\\
&=&\delta(1-1')
+\int\limits_{{\cal C}}d{\bar 1}\left[\Sigma^c(1{\bar 1})+
\Sigma^{in}(1{\bar 1})\right]g_1({\bar 1}1'),
\end{eqnarray}
from which we obtain the KBE for the correlation functions, if we
restrict the time arguments to opposite branches of the contour:
\begin{eqnarray}\label{KBG g><}
\left(i\frac{\partial}{\partial t_1}+\frac{\nabla_1^2}{2m}\right)
g_1^{\gtrless}(11')-\int d{\bar 1}\,U(1{\bar 1})
g_1^{\gtrless}({\bar 1}1')
-\int d{\bar r}_1\,\Sigma^{HF}(1{\bar 1})g_1^{\gtrless}({\bar 1}1')
\nonumber\\
=\int\limits_{t_0}^{\infty}d{\bar 1}\,
\Sigma^R(1{\bar 1})g_1^{\gtrless}({\bar 1}1')
+\int\limits_{t_0}^{\infty}d{\bar 1}\left[\Sigma^{\gtrless}(1{\bar 1})+
\Sigma^{in}(1{\bar 1})\right]g_1^A({\bar 1}1').
\end{eqnarray}
In contrast to the original KBE, there are two important new
properties which have to be underlined here: The equations (\ref{KBG g><})
are valid for an {\em arbitrary initial time point} $t_0$,
and they explicitly contain the influence of {\em arbitrary initial
correlations} in the additional self--energy term $\Sigma^{in}$.
The analytical properties of the retarded and advanced Green's functions
give rise to a damping $\gamma_{12}$ leading to a decay of the initial
correlation term after a  time of the order $t\sim 1/\gamma_{12}\sim\tau_{cor}$.
Thus, there is no need at all to postulate Bogolyubov's weakening condition;
for $t>\tau_{cor}$, the generalized Kadanoff--Baym equations switch
from the initial regime into the kinetic, or Bogolyubov regime,
``automatically''.

\section{Non--Markovian kinetic equation in binary collision approximation}
In order to compare our results with previous work in the field of kinetic
theory, we will derive a non--Markovian kinetic equation from the generalized
KBE. To get such an equation for the single--time Wigner distribution, the
following steps are necessary:
\begin{itemize}
\item The KBE (\ref{KBG g><}) and its adjoint have to be taken in the equal time
limit and substracted from each other to yield the time--diagonal KBE.
\item Since this equation is not yet a closed equation for the Wigner function,
the reconstruction problem $g^{\gtrless}=g^{\gtrless}[f]$ has to be solved.
We use the generalized Kadanoff--Baym ansatz proposed by Lipavsk\'{y} et al.
\cite{lipavski-etal.86}.
\item An approximation for the self--energy has to be introduced. Here, we
take $\Sigma$ in binary collision (T--Matrix) approximation.
\end{itemize}
Performing these three steps, we get the non--Markovian Boltzmann equation
\cite{KBKS97}:
\begin{eqnarray}\label{BE}
&&\left( \frac{\partial }{\partial T}+\frac{P}{m}\nabla _{R}\right)
f_{1}(T)=I^{IC}(T)+{\rm Tr}\int\limits_{t_{0}}^{T}d\bar{t}\;d\stackrel{=}{t}
\;d\stackrel{\equiv }{t} \nonumber\\
&\times& \left\{T^{R}(T,\stackrel{=}{t})\;\bar{U}^{A}(\stackrel{=}{t},
\stackrel{\equiv }{t})\;T^{A}(\stackrel{\equiv }{t},\bar{t})\;U^{A}(\bar{t},T)
\;\left[ F^{<}(\bar{t})\bar{F}^{>}(\stackrel{\equiv }{t})-\bar{F}^{>}(\bar{t})
\bar{F}^{<}(\stackrel{\equiv }{t})\right]\right. \nonumber\\
&&+ \left.T^{R}(T,\stackrel{=}{t})\;U^{R}(\stackrel{=}{t},\stackrel{\equiv }{t})
\;T^{A}(\stackrel{\equiv }{t},\bar{t})\;U^{A}(\bar{t},T)\;\left[ F^{<}(\bar{t})
\bar{F}^{>}(\stackrel{=}{t})-F^{>}(\bar{t})\bar{F}^{<}(\stackrel{=}{t})\right]
 \right\}\nonumber\\
&&+adjoint
\end{eqnarray}
with
\begin{eqnarray*}
F^{<}&=&f_{1}^{<}\;f_{2}^{<}\;=f_1\;f_2\;,\quad
F^{>}=f_{1}^{>}\;f_{2}^{>}=\left( 1\pm f_{1}\right) \;\left( 1\pm f_{2}\right)\\
\stackrel{(-)}{U}^{R/A}(tt^{\prime })&=&\Theta [\pm (t-t^{\prime })]\;
e^{i\stackrel{(-)}{E}_{12}(t-t^{\prime})}
\end{eqnarray*}
The initial correlation contribution reads  
\begin{eqnarray}\label{IIC}
&&I^{IC}(t)=n\,{\rm Tr}\left\{\int d\bar{t}\;T^{R}(t,\bar{t})\;U^{0^R}(\bar{t}
,t_{0})\;c(t_{0})\;U^{0^A}(t_{0},t)\right. \nonumber\\
&&\left.-\int d\bar{t}d\stackrel{=}{t}d\stackrel{\equiv }{t}\;
g_{12}^{0^R}(t,\bar{t})\;
T^{R}(\bar{t},\stackrel{=}{t})\;U^{0^R}(\stackrel{=}{t},t_{0})\;c(t_{0})\;
U^{0^A}(t_{0},\stackrel{\equiv }{t})\;T^{A}(\stackrel{\equiv }{t},t)\right\}
\nonumber\\
&&+adjoint
\end{eqnarray}
Here, $T^{R/A}(t,t')$ is the off--shell two--particle scattering matrix given
by a generalized Lippmann--Schwinger equation.

\section{Numerical illustrations and discussion}
For illustration of our theoretical results, we have performed
numerical solutions of the Kadanoff--Baym equations including
initial correlations. We considered the
relaxation of electrons in dense hydrogen with self--energies in
second Born approximation. Starting from an initial nonequilibrium
distribution, we compared the relaxation for two cases:
I) without initial correlations, $C(t_0)\equiv 0$ and II), with nonzero initial
correlations which were chosen in the form of the Debye pair correlation
function
\begin{eqnarray}
C(q,p_1,p_2,t_0)= - \frac{V_D(q)}{kT} f(p_1) f(p_2)[1-f(p_1+q)][1-f(p_2-q)],
\label{g-debye}
\end{eqnarray}
where $f \equiv f(t_0)$.
As expected, the presence of initial correlations turns out to be
important on short times. This becomes particularly clear from analyzing
the time evolution of the mean potential and kinetic energy, Fig. 1.
While for the uncorrelated initial state, potential energy starts with
zero and builds up continuously, the picture changes if there exist
initial correlations. With the choice of the form (\ref{g-debye}),
the correlations are stronger than in equilibrium which corresponds to
a larger magnitude of potential energy at $t=t_0$, which, consequently,
is reduced in the course of the relaxation. Due to conservation of total
energy, kinetic energy shows exactly the opposite trend.
One clearly sees the decay of the initial correlation term, as the curves
for the two cases merge after times of the order of the correlation
time. This confirms that indeed, Bogolyubov's weakening
principle is reproduced by the presented generalized Kadanoff--Baym equations
in a dynamic and selfconsistent way.

\begin{figure}
\psfig{figure=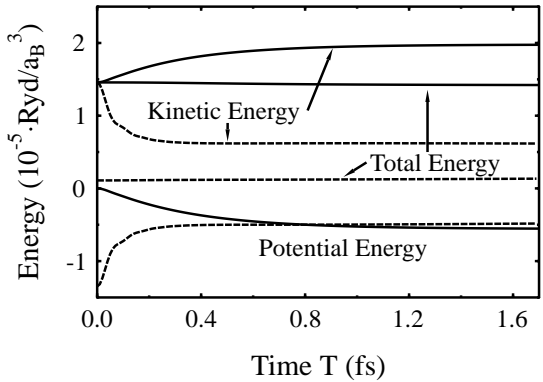,width=9cm,angle=0}
\caption{\label{4} Time evolution of kinetic, potential, and total energy for
zero (solid lines) and nonzero (dashed lines) initial correlations. The initial
distribution is an uncorrelated equilibrium (Fermi) distribution with
$T=10000 K$ and $n=10^{21} cm^{-3}$.}
\end{figure}

Our numerical results illustrate the effect of initial correlations on 
the short--time relaxation behavior for a simple model case. But our
theoretical approach is completely general and allows for numerical
investigations of far more complex initial correlations. Besides the fundamental
interest in the problem of initial correlations in the Kadanoff--Baym equations,
our results are also of practical importance. With the possibility to 
start quantum kinetic calculations from a general initial state, 
the scope of nonequilibrium processes in many--body systems which are 
accessible for numerical investigation is essentially 
extended. Although the determination of $C(t_0)$ can be complicated by itself, 
our approach allows to separate this problem (the ``generation'' of the 
correlated state) from the relaxation dynamics.

\section*{Acknowledgments}
This work is supported by the Deutsche
Forschungsgemeinschaft (SFB 198 and Schwerpunkt ``Quantenkoh\"arenz in
Halbleitern'').

\end{document}